\documentclass[twocolumn,superscriptaddress,amsmath,amssymb,aps,pra]{revtex4}
\usepackage{amsmath,amsthm,amssymb}
\usepackage{latexsym}
\usepackage{amscd,graphicx}
\usepackage[titletoc]{appendix}
\usepackage{epsfig}

\usepackage{epstopdf}
\usepackage[normalem]{ulem}
\usepackage[dvipsnames]{xcolor}
\usepackage[colorlinks=true,linkcolor=blue,citecolor=black]{hyperref}

\usepackage{braket}

\def \ada {{a}^\dagger a}
\def \a   {{a}}

\def \ad  {{a}^\dagger}
\def \add {{a}^{\dagger2}}

\def \sx  {{\sigma}_x}

\def \sz  {{\sigma}_z}

\def \H   {{H}}

\begin{document}
	
	\title{Switchable Superradiant Phase Transition with Kerr Magnons}
	
	\author{Gang Liu}
	\affiliation{School of Physical Science and Technology, Lanzhou University, Lanzhou 730000, China}
	\affiliation{Key Laboratory for Quantum Theory and Applications of MoE, Lanzhou Center for Theoretical Physics, and Key Laboratory of Theoretical Physics of Gansu Province, Lanzhou University, Lanzhou, Gansu 730000, China}
	
	\author{Wei Xiong}
	\altaffiliation{xiongweiphys@wzu.edu.cn}
	\affiliation{Department of Physics, Wenzhou University, Zhejiang 325035, China}
	
	\author{Zu-Jian Ying}
	\altaffiliation{yingzj@lzu.edu.cn}
	\affiliation{School of Physical Science and Technology, Lanzhou University, Lanzhou 730000, China}
	\affiliation{Key Laboratory for Quantum Theory and Applications of MoE, Lanzhou Center for Theoretical Physics, and Key Laboratory of Theoretical Physics of Gansu Province, Lanzhou University, Lanzhou, Gansu 730000, China}
	
	\date{\today}
	
	\begin{abstract}
		The superradiant phase transition (SPT) has been widely studied in cavity quantum electrodynamics (CQED). However, this SPT is still subject of ongoing debates due to the no-go theorem induced by the so-called ${\bf A}^2$ term (AT). We propose a hybrid quantum system, consisting of a single-mode cavity simultaneously coupled to both a two-level system and yttrium-iron-garnet sphere supporting magnons with Kerr nonlinearity, to restore the SPT against the AT. The Kerr magnons here can effectively introduce 
an additional AT tunable and strong enough
to counteract the intrinsic AT, via adiabatically eliminating the degrees of freedom of the magnons. We show that the Kerr magnons induced SPT can exist in both cases of ignoring and including the intrinsic AT. Without the intrinsic AT, the critical coupling strength can be dramatically reduced by introducing the Kerr magnons, which greatly relaxes the experimental conditions for observing the SPT. With the intrinsic AT, the forbidden SPT can be recovered with the Kerr magnons in a reversed way. Our work paves a potential way to manipulate the SPT against the AT in hybrid systems combining CQED and nonlinear magnonics.
		
	\end{abstract}
	
	\pacs{Valid PACS appear here}
	\keywords{Suggested keywords}
	
	\maketitle
	
	\section{Introduction}
	
	With the experimental advances into the era of ultra-strong coupling in light-matter interactions~\cite{Diaz2019RevModPhy,Kockum2019NRP} and the theoretical efforts on the fundamental Quantum Rabi model (QRM)~\cite{Braak2011,Boite2020,Liu2021AQT,
		Irish-class-quan-corresp,Irish2017,Ashhab2013,Ying2015,Hwang2015PRL,Ying2020-nonlinear-bias,
		Ying-2021-AQT,Ying-gapped-top,Ying-Stark-top,Ying-Winding,LiuM2017PRL,Hwang2016PRL,Ying-2018-arxiv,CongLei1719,
		Grimaudo2023-PRL,Grimaudo2023-Entropy,ChenQH2012,Yan2023-AQT,PengJie2019,
		Padilla2022,Garbe2020,Garbe2021-Metrology,Ying2022-Metrology,PRX-Xie-Anistropy,
		Batchelor2015,Batchelor2021,ChenGang2012,Eckle-2017JPA,Casanova2018npj,Braak2019Symmetry,Minganti-2023-SPT-3Body,
		Ma2020Nonlinear,Gao2021,Stark-Cong2020,Cong2022Peter,Zheng2017,
		Rabi-Braak,Eckle-Book-Models,JC-Larson2021}, finite-component quantum phase transitions (QPTs) have recently received an increasing attention~\cite{Liu2021AQT,Ashhab2013,Ying2015,Hwang2015PRL,Ying2020-nonlinear-bias,Ying-2021-AQT,Ying-gapped-top,Ying-Stark-top,Ying-Winding,LiuM2017PRL,
		Hwang2016PRL,Ying-2018-arxiv,Grimaudo2023-PRL,Grimaudo2023-Entropy,Irish2017,Minganti-2023-SPT-3Body,CongLei1719}. Practical applications of the finite-component QPTs have been exploited in critical quantum metrology and quantum information science~\cite{Garbe2020,Garbe2021-Metrology,Ilias2022-Metrology,Ying2022-Metrology,Cai-2021,Wang-2018,Zhu-2019}.

	A QPT occurs in the ground state in the variation of some non-thermal parameter and is regarded to be driven by quantum fluctuations~\cite{Sachdev-QPT,Irish2017}. Besides topological transitions under symmetry protection~\cite{Ying-2021-AQT,Ying-gapped-top,Ying-Stark-top,Ying-Winding} and various transitions in symmetry breaking~\cite{Ying2020-nonlinear-bias,Ying-2018-arxiv}, the superradiant phase transition (SPT) is a typical QPT in light-matter interactions.
	The SPT was first predicted in the Dicke model~\cite{Hepp-1973-1,Hepp-1973-2} consisting of an ensemble of $N$ two-level systems (TLSs) coupled to photons in a cavity~\cite{Dicke-1954}. By varying the coupling strength, the ground state of the system changes abruptly from the normal phase (NP) to the supperadiant phase (SP) with a boost of photon number. Due to this exotic behavior, the SPT has been widely
	studied~\cite{Liu2021AQT,Ashhab2013,Ying2015,Hwang2015PRL,Ying2020-nonlinear-bias,Ying-2021-AQT,LiuM2017PRL,
		Hwang2016PRL,Ying-2018-arxiv,Ying-Winding,Grimaudo2023-PRL,Grimaudo2023-Entropy,Irish2017,
		Li-2006,Baumann-2011,Baksic-2014,Zou-2014,Bamba-2016,Kirton-2017,Wangym-2020,Minganti-2023-SPT-3Body,CongLei1719}. Besides the Dicke model, the QRM~\cite{Rabi-1936,Rabi-1937,Rabi-Braak,Eckle-Book-Models,JC-Larson2021}, describing the interaction between a single TLS (with level splitting $\Omega$) and a single-mode field (with frequency $\omega$), can also predict the SPT in the low-frequency limit~(i.e., $\omega/\Omega\rightarrow 0$) as a replacement of thermodynamic limit~\cite{Liu2021AQT,Ashhab2013,Ying2015,Hwang2015PRL,Ying2020-nonlinear-bias,Ying-2021-AQT,LiuM2017PRL,
		Hwang2016PRL,Ying-2018-arxiv,Ying-Winding,Grimaudo2023-PRL,Grimaudo2023-Entropy,Irish2017,Minganti-2023-SPT-3Body,
		Bakemeier-2012,CongLei1719}. Both the Dicke model and the QRM can be experimentally realized in cavity quantum electrodynamics (CQED) or circuit systems~\cite{Niemczyk-2010,Anappara-2009,Mlynek-2014,Breeze-2017,Braumuller-2017}. However, the no-go theorem induced by the so-called $\text{\bf A}^2$ term (AT) (i.e., the squared electromagnetic vector potential) in realistic CQED forbids the SPT~\cite{Rzazewski-1975}.
 Actually whether or not the SPT is prohibited by the AT has been a subject of much debate for several decades~\cite{
 Rzazewski-1975,Birula-1979,
 Knight-1978,Keeling-2007,Viehmann-2011,Ciuti-2012-comm,Ciuti-2010,Vukics-2012,Liberato-2014,Vukics-2014,Hartmann-2014,Jaako-2016,
 Andolina-2019,Andolina-2020,Andolina-2022,
 Adam-2022,Adam-2020,Adam-2019,
 Champel-2019,Stefano-2019,Bamba-2016-PRL,Bamba-2017-PRA}. Indeed, both no-go theorems~\cite{Rzazewski-1975,Birula-1979,Andolina-2019,Andolina-2020,Andolina-2022} and counter no-go theorems~\cite{Adam-2022,Adam-2020,Adam-2019,Ciuti-2010,Vukics-2012} have been raised, essentially depending on different situations such as adopting the conventional two-level approximation (qubit) 
 or non-truncated Hilbert space for the matter part~\cite{Rzazewski-1975,Birula-1979,Andolina-2019,Andolina-2020,Andolina-2022,Stefano-2019,Adam-2019},
 applying the minimal replacement rule of gauge to kinetic terms
 or also to non-local potentials 
 ~\cite{Stefano-2019,Adam-2022,Adam-2020,Adam-2019}, considering spatially uniform cavity field~\cite{Rzazewski-1975,Birula-1979} or spatially varying electromagnetic field~\cite{Andolina-2020,Andolina-2022,Champel-2019}. An arbitrary-gauge approach suggests that the conflicting no-go and counter no-go theorems may be reconciled in many-dipole cavity QED
systems as views of different quantum
subsystems~\cite{Adam-2022,Adam-2020}.
Besides natural atomic systems, the debate has been extended to artificial atomic systems as well~\cite{Ciuti-2010,Viehmann-2011,Ciuti-2012-comm,Jaako-2016,Bamba-2016,Adam-2022}, as in circuit systems the existence of an equivalent AT is also controversial
        and depends on specific circuit designs~\cite{Jaako-2016,Bamba-2016-PRL,Bamba-2017-PRA}.
	
	Other effort to circumvent the difficulty of reaching a consensus on the existence of the AT is to find possibility to cancel the AT~\cite{Chen-2021-NC,Lu-2018-1,Lu-2018-2}. It has been suggested that the disappearing SPT of the Rabi~(Dicke) model in the presence of the AT can be regained by combining the optomechanics and CQED~\cite{Lu-2018-1,Lu-2018-2}, where the customarily-used Coulomb gauge is chosen under two-level approximation. The optomechanics there actually provides an auxiliary AT at the single-photon level to compensate the intrinsic AT. However, the required strong quadratical optomechanical coupling at the single-photon level is still a challenge within current fabrication techniques~\cite{Aspelmeyer-2014}.  In such a situation, an alternative quantum system capable of introducing strong and tunable auxiliary AT is highly desirable, although the arbitrary-gauge approach~\cite{Adam-2022,Adam-2020,Adam-2019} has been proposed to recover the SPT in CQED systems with the AT.
	
	In this regard, a feasible way may lie in nonlinear cavity magnonics. With the advancement of quantum materials, magnons (i.e., quanta of spin wave) in a yttrium-iron-garnet (YIG) sphere with flexible controllability and high spin density have received much attention theoretically and experimentally~\cite{Rameshti-2022,Wang-2020,Yuan-2022,Quirion-2019}, especially in magnon dark modes~\cite{Zhang-2015}, spin currents~\cite{Bai-2015,Bai-2017}, entanglement~\cite{Li-2018,Yuan-2020,Sun-2021,Zhanggq-2022,Qi-2022}, nonreciprocity~\cite{Wangym-2022}, and microwave-to-optical transduction~\cite{Hisatomi-2016,Zhu-2020}. Moreover, the magnons in the YIG sphere can have tunable Kerr nonlinearity via controlling the external magnetic field, originating from the magnetocrystalline anisotropy~\cite{Stancil-2009,Zhanggq-2019}. This nonlinearity has been demonstrated in experiment~\cite{Wangyp-2016} and used to study the bi- and multistabilities~\cite{Wangyp-2018,nair-2020,shen-2021}, quantum entanglement~\cite{zhangz-2019}, quantum phase transition~\cite{zhang-2021}, long-range spin-spin coupling~\cite{Xiong-2022,Tian-2023}, and nonreciprocal entanglement in cavity-magnon optomechanics~\cite{Chen-2023}.
	
	In the present work, we open a possible avenue to gain 
a
strong and tunable
auxiliary AT which is capable of switching on or manipulating the SPT, either the AT is present or not as in the
afore-mentioned debate on the no-go theorem.
Based on the advances in magnons, 
we
propose a hybrid system consisiting of a microwave cavity simultaneously coupled to both a TLS and Kerr magnons (i.e., magnons with the Kerr nonlinearity) in a YIG sphere to restore the SPT. 
Here the coupled-TLS-cavity subsystem and the coupled-magnon-cavity subsystem build the Rabi model and cavity magnonics, respectively. By adiabatically eliminating the degrees of freedom of the magnons, we demonstrate that the Kerr magnons finally play a role to effectively introduce an additional AT which can counteract the original one. Owing to 
the experimental controlability and the nonlinearity-enhanced effect,
the additional AT can be 
tunable and strong enough
to switch on and manipulate the SPT. In the absence of the AT, the critical coupling of the SPT can be significantly reduced when the Kerr magnons are included, while in the presence of the AT the vanishing SPT can be restored by the Kerr magnons. Our proposal shows that the combination of CQED and nonlinear cavity magnonics can provide a potential platform to study quantum critical physics.
	
	\section{Model of hybrid system}
	
\begin{figure}
		\centering
		\includegraphics[width=0.85\linewidth]{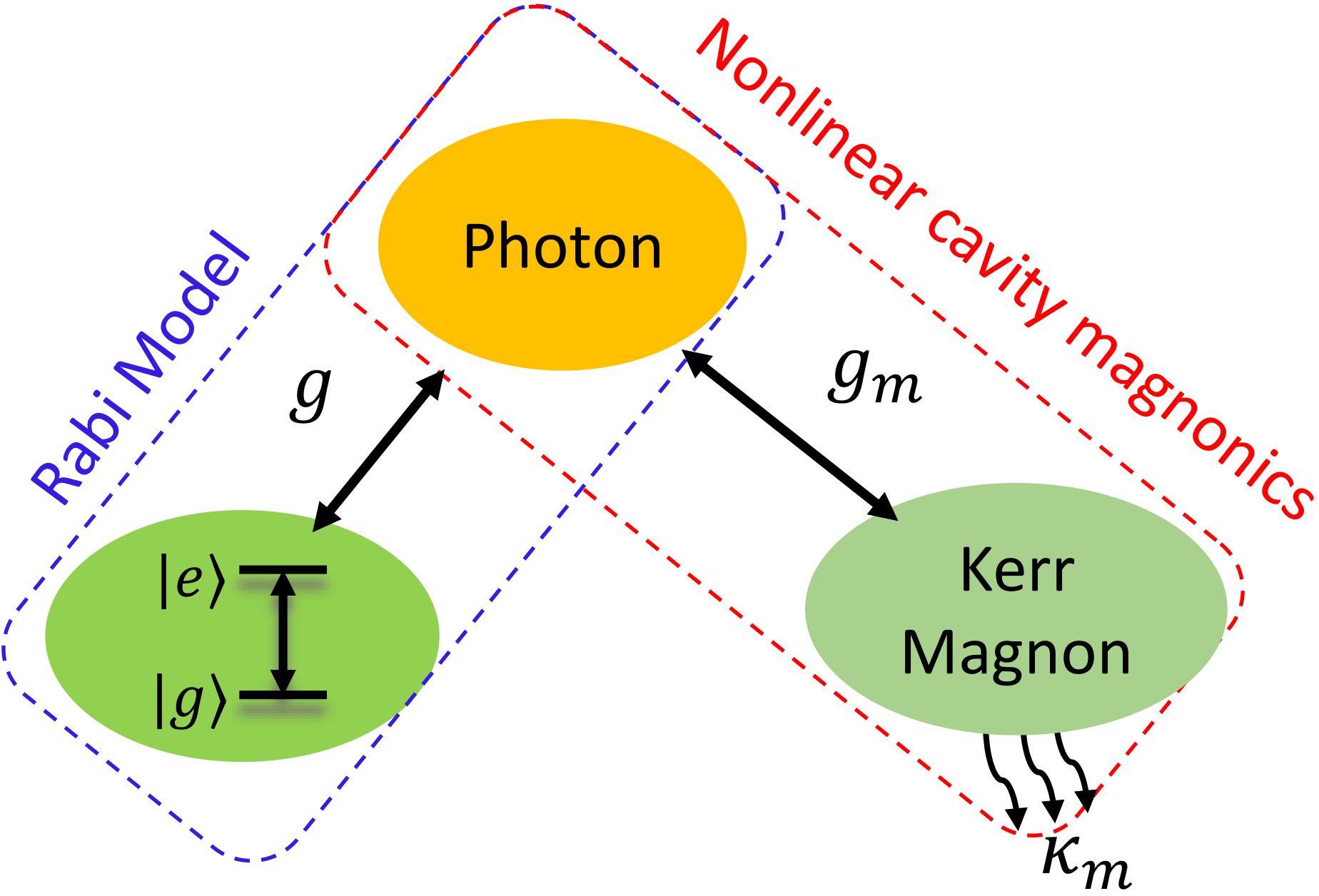}
		\caption{Sketch of a hybrid quantum model including a Rabi model coupled to a Kerr magnon mode. Here $g$ ($g_m$) is the Rabi (magnon) coupling strength of the two atomic levels (the magnon mode), respectively. Here the cavity mode is represented by photon and $\kappa_m$ is the magnon decay rate. 
		}
		\label{fig:cavity}
	\end{figure}

	We consider a hydrid quantum system consisting of
	a microwave cavity~\cite{Wangyp-2016} or superconducting resonator~\cite{Wangyp-2019})
	simultaneously coupled to a TLS, such as superconducting qubits~\cite{You-2011,Xiang-2013}, and Kerr magnons in a YIG sphere~(see Fig.~\ref{fig:cavity}), where the Kerr nonlinearity of the magnons stems from the magnetocrystalline anisotropy. The Hamiltonian of the proposed hybrid system can be written as (setting $\hbar=1$)
	\begin{align}
		\H=\H_{\rm Rabi}+\H_{A^2}+\H_K+\H_I,\label{eq:totalHamiltonian}
	\end{align}
	where $H_{\rm Rabi}=\omega \ada + \frac{1}{2}\Omega {\hat \sigma}_z + g\sx(\a + \ad)$ 
is the QRM, describing the interaction between the TLS and the cavity. The parameter $\omega$ ($\Omega$) is the frequency of the cavity (the TLS) and $g$ is the coupling strength. The operators
$\sx = \ket{e}\bra{g} + \ket{g}\bra{e}$ and $\sz = \ket{e}\bra{e} - \ket{g}\bra{g}$ represent the Pauli matrices of the TLS, while $a$ and $a^\dag$ denote the annihilation and creation operators of the cavity, respectively. The second term $\H_{A^2}= ({\alpha g^2}/{\Omega})(\a + \ad)^2$ is the AT, where $\alpha\ge 1$ for the achieved Rabi model (Dicke model) in cavity quantum electrodynamics, is dependent on the Thomas-Reiche-Kuhn sum rule \cite{Ciuti-2010}. The Kerr Hamitonian $\H_K=\omega_m m^\dag m + Km^\dag mm^\dag m$, with the frequency $\omega_m=\gamma B_0+2\mu_0K_{\rm an}\gamma^2/(M^2V_m^2)-2\mu_0\rho_s sK_{\rm an}\gamma^2/M^2$ and the Kerr coefficient $K/\hbar=2\mu_0K_{\rm an}\gamma^2/(M^2V_m^2)$, represents the interaction among magnons and provides the anharmonicity of the magnons. Here, $\gamma/2\pi=g_e\mu_B/\hbar$ is the gyromagnetic ration with the $g$-factor $g_e$ and the Bohr magneton $\mu_B$,
	$s=\hbar/2$ is the spin quantum number,
	$\rho_s=2.1\times10^{-22}~{\rm cm}^{-3}$ is the spin density of the YIG sphere, $\mu_0$ is the vacuum permeability, $K_{\rm an}$ is the first-order anisotropy constant of the YIG sphere, $B_0$ is the amplitude of a bias magnetic field along $z$-direction, $M$ is the saturation magnetization, and $V_m$ is the volume of the YIG sphere. Note that the Kerr coefficient $K$ can be either positive or negative by tuning the angle between
	crystallographic axis [100] or [110] of the YIG sphere and the bias magnetic field~\cite{shen-2021,Zhanggq-2019,Wangyp-2016}. The last term $\H_I=g_m (\a m^\dag + \ad m)$ characterizes the interaction between the photons in the cavity and the magnons in the YIG sphere with coupling strength $g_m$.

\section{Effective Hamiltonian of the hybrid system}

Taking account of the dissipations of the considered system in Eq.~(\ref{eq:totalHamiltonian}), its dynamics is governed by the Heisenberg-Langevin equation~\cite{Benguria-1981}, i.e.,
	\begin{align}
		\frac{d \mathcal{O}}{d t} = -i[\mathcal{O},H] + \mathcal{L}^\dagger \mathcal{O} \mathcal{L} -\frac12( \mathcal{O} \mathcal{L}^\dagger \mathcal{L}+ \mathcal{L}^\dagger \mathcal{L} \mathcal{O}),
		\label{eq:HLE}
	\end{align}
	where $\mathcal{O}$ represents the system operator and $\mathcal{L}$ is the Lindblad operator. For the magnon-cavity subsystem of interest,  the relaxation operators can be defined as $\mathcal{L}_{a} = \sqrt{\kappa} \a$ and $\mathcal{L}_m = \sqrt{\kappa_m}  m$, where $\kappa$ ($\kappa_m$) is the decay rate of the cavity (magnon). Specifically, the dynamics of the magnon-cavity subsystem can be written as
	\begin{align}
		\dot{\a}(t)	=& -(\frac\kappa2+i\omega) \a -ig\sx-ig_m  m\notag\\
		& -2i ({\alpha g^2}/{\Omega})(\a + \ad)+ \sqrt{\kappa}\a_{\rm in},\notag\\
		\dot{ m}(t) =& -(\frac{\kappa_m} {2}+i\omega_m)  m - 2i Km^\dag m  m \label{eq3}\\
		&-ig_m \a +  \sqrt{\kappa_m} m_{\rm in},\notag
	\end{align}
	where $\a_{\rm in}$ and $ m_{\rm in}$ are the vacuum input noise operators of the cavity and magnon, respectively. The corresponding mean values are zero, i.e., $\langle\a_{\rm in}\rangle = \langle m_{\rm in}\rangle =0$.  By rewritting each operator of the magnon-cavity susbsystem as its expectation value plus the corresponding fluctuation, i.e., $a\rightarrow \langle a\rangle+a$ and $m\rightarrow \langle m\rangle+m$, the nonlinear term (i.e., $m^\dag m m$) in Eq.~(\ref{eq3}) can be re-expressed as
	\begin{align}
		\begin{split}
			m^\dag  m   m  &\rightarrow |\langle{ m}\rangle|^2\langle m\rangle+2|\langle{ m}\rangle|^2 m+\langle m^\dag\rangle  m^2\\
			+& \langle m\rangle^2 m^\dag+ 2\langle m\rangle m^\dag m+m^\dag  m^2.
		\end{split}
		\label{eq4}
	\end{align}
	By neglecting the high-order fluctuation terms, we can linearize the dynamics in Eq.~(\ref{eq3}) as
	\begin{align}
		\dot{\a}(t)	=& -(\frac\kappa2+i\omega) \a -ig\sx-ig_m  m\notag\\
		& -2i ({\alpha g^2}/{\Omega})(\a + \ad)+ \sqrt{\kappa}\a_{\rm in},\notag\\
		\dot{ m}(t) =& - (\frac{\kappa_m} {2}+i \Delta_m) m   -2i K\langle{ m}\rangle^2 m^\dagger \label{eq:dya}\\
		&-ig_m \a +  \sqrt{\kappa_m} m_{\rm in},\notag
	\end{align}
	where $ \Delta_m = \omega_m + 4 K  N_m$, with the mean magnon number $ N_m = |\langle m\rangle|^2$, is the effective frequency of the magnon induced by its Kerr nonlinearity.
	
	Note the magon-cavity coupling $g_m$ can be tuned by the displacement of the YIG sphere \cite{Tune-gm-NC2017}, we can assume a weak coupling $\kappa_m\gg g_m$ so that the decoherence time of magnons is much shorter than that of photons in the cavity. In such a situation, the degrees of freedom of the magnon can be adiabatically eliminated by setting $\dot{ m}(t)=0$, which directly gives rise to
	\begin{align}
		m =  \frac{2 K\langle{ m}\rangle^2 g_m}{W}\ad - \frac{g_m( \Delta_m + i \frac{\kappa_m} {2})}{W}\a,	\label{eq:s}
	\end{align}
	where $W=\Delta_m^2 + \frac{\kappa_m^2} {4} - 4 K^2 N_m^2$.
	Substitution of Eq. \eqref{eq:s} back into Eq. \eqref{eq:dya} leads us to
	\begin{align}
		\dot{\a}
		=&  -[\frac{\kappa_{\rm eff}}2+i(\omega-\eta \Delta_m)]\a -ig\sx	\label{eq7}\\
		& 	-2i({\alpha g^2}/{\Omega})(\a + \ad)-2i K\eta\langle{ m}\rangle^2 \ad + \sqrt{\kappa_{\rm eff}}\a_{\rm in},\notag
	\end{align}
	where $\eta = { g_m^2}/W$ is a dimensionless parameter related to the  cavity frequency shift $\eta \Delta_m$,  $\kappa_{\rm eff} = \kappa +\kappa_m\eta$ is the effective decay rate of the cavity. By rewriting the equation of motion in Eq.~(\ref{eq7}) as $\dot{\a}=-i[\a,H_{\rm eff}]- ({\kappa_{\rm eff}}/2) \a+\sqrt{\kappa_{\rm eff}}\a_{\rm in}$, we obtain the effective Hamiltonian after eliminating the degrees of freedom of the Kerr magnons
	\begin{align}
		H_{\rm eff}=\H_{\rm Rabi}+\H_{A^2} + K\eta \langle m\rangle^2 ( \ad - \frac{ \Delta_m }{2K \langle m\rangle^2}\a  )^2.\label{eq8}
	\end{align}
	As $\Delta_m$, $K$ and $\langle m\rangle$ can be tuned by the amplitude and the direction of the bias field,
	the YIG sphere size and the drive power, with the sign of $K$ also being adjustable~\cite{shen-2021,Zhanggq-2019,Wangyp-2016}, we can assume $K<0$ and set $ \Delta_m = -2K \langle m\rangle^2$. Thus, Hamiltonian~(\ref{eq8}) reduces to
	\begin{align}
		H_{\rm eff}=\H_{\rm Rabi}+\H_{A^2}+	\H_{A^2}^\prime,\label{eq9}
	\end{align}
	where $\H_{A^2}^\prime=-\chi(\a+\ad)^2$, with $\chi\equiv - K\eta\langle m\rangle^2$ and reduced $\eta = { 4g_m^2}/\kappa _m^2$, is additional AT induced by the Kerr magnons. One sees that the strength of the additional AT $\chi$ is highly enhanced by the nonlinear Kerr effect as in proportion to $\langle m\rangle^2$, thus being competent to counteract the effect of the original AT $H_{A^2}$ and restore the SPT of the QRM in the presence of the AT.
	
	\begin{figure}
		\includegraphics[width=1.0\columnwidth]{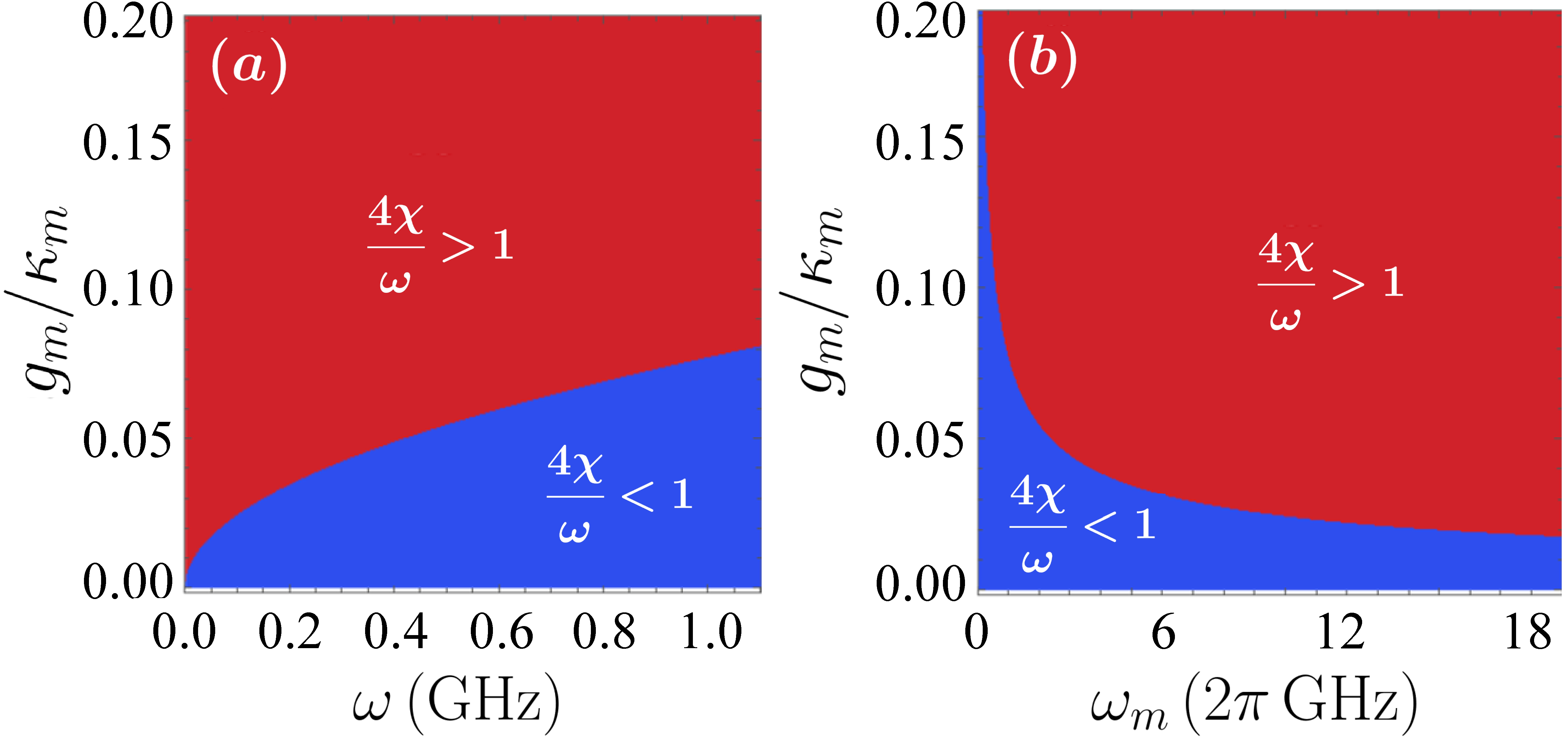}
		\caption{SPT-restored regime (red, marked by $4\chi/\omega>1$) in $\omega$-$g_m$ plane (a) at $\omega _m/2\pi=10$ GHz and $\omega _m$-$g_m$  plane (b) at $\omega=0.1$ GHz.   }
		\label{fig:parameters}
	\end{figure}

\section{Switchable SPT by Kerr magnonic coupling}

To study the ground-state properties of the Hamiltonian in Eq.~(\ref{eq9}), a squeezing transformation, i.e., $ S(r) = \exp[r(\add - \a^2)/2]$, is imposed, leading to $\a \rightarrow \a \cosh(r) + \ad\sinh(r)$. By choosing the squeezing parameter
	\begin{align}
		r = -\frac14\ln\left(1 + \alpha \overline{g}^2 - 4\frac{\chi}{\omega}\right),
		\label{eq:nogo}
	\end{align}
	with rescaled coupling $\overline{g} = g /g_c$ where $g_c=\sqrt{\omega\Omega}/2$ is the critical coupling of the QRM without the AT~\cite{Ashhab2013,Ying2015,Ying2020-nonlinear-bias}, Hamiltonian (\ref{eq9}) is transformed to an effective QRM
	\begin{align}\label{eq:likeQRM}
		\H_S=U^\dag H_{\rm eff}U = \tilde{\omega} a^\dagger a + \frac{\Omega}{2}\sigma_z + \tilde{g} (a^\dagger+a)\sigma_x,
	\end{align}
	where $\tilde{\omega} = \omega e^{-2r}$ is the renormalized photon frequency in the cavity, $\tilde{g} = g e^r$ is the effective coupling strength. Compared to the original QRM, the parameters $\tilde{\omega}$ and $\tilde{g}$ in Eq.~\eqref{eq:likeQRM} are tunable via $\chi$. Correspondingly, with renormalized critical coupling $\tilde{g}_c=\sqrt{\tilde{\omega}\Omega}/2$, the rescaled coupling strength in Eq.~(\ref{eq:likeQRM}) becomes
	\begin{align}
		\tilde{\overline{g}} \equiv \tilde{g} /\tilde{g}_c =\overline{g}/\sqrt{1 + \alpha \overline{g}^2 - 4\chi/\omega}.\label{eq12}
	\end{align}
	The critical point is decided by $\tilde{\overline{g}} _c= 1$, beyond which the ground state transits from the NP to the SP.
	
%
	
	Obviously, when Kerr magnons are not included (i.e., $K=0$ or equivalently $\chi=0$), $\tilde{\overline{g}}$ reduces to $\tilde{\overline{g}} \rightarrow \overline{g}/\sqrt{1 + \alpha \overline{g}^2}$. For the realistic cavity QED, $\alpha\ge 1$ leads to $\tilde{\overline{g}} < 1$, indicating that the Rabi model with the AT is always in the normal phase, unable to reach the SPT. This is just the afore-mentioned no-go theorem~\cite{Ciuti-2010}.
	
	When we have Kerr magnons, the critical coupling is explicitly determined by $\overline{g} _c = \sqrt{(1- 4\chi/\omega)/(1-\alpha)}$ in terms of the original system parameters. This indicates that the SPT is switchable via the Kerr magnons as long as $\chi/\omega>1/4$ for $\alpha>1$, which is equivalent to $g_m/\kappa_m>\sqrt{3\omega/8\omega_m}$ by the constraint $ \Delta_m = -2K \langle m\rangle^2$. Fig. \ref{fig:parameters} illustrates the SPT-restored regime (red, marked by $4\chi/\omega>1$). These conditions can be readily fulfilled experimentally~\cite{Wangyp-2018,Tune-gm-NC2017,shen-2021,Wangyp-2016,Blais-2004} as $|K|=0\sim 10^3$ nHz, $\langle m \rangle=0\sim 10^{17}$, with a typical order $2\pi\times10$ GHz for $\omega _m$ and the low frequency requirement of $\omega$ for SPT, while $g_m$ can be tuned from weak couplings ($g_m/\kappa_m<1$) to strong couplings ($g_m/\kappa_m>1$) by the YIG-sphere displacement~\cite{Tune-gm-NC2017}.

	\section{Reversed SPT and phase diagram}
	
	In the absence of the AT, the SPT occurs in increasing the coupling strength as small-$g$ regime is the NP. However, in the presence of the AT, the SPT restored by Kerr magnons is reversed. In fact, the ground state is in the SP for regime $\tilde{\overline{g}} > 1$ which gives rise to $\overline{g} < \overline{g} _c $. This means the SP lies below $\overline{g} _c$ instead of above $\overline{g} _c$. Reversely, the regime $\overline{g} > \overline{g} _c $ is in the NP, as derived from $\tilde{\overline{g}} < 1$. The reversed SPT would facilitate experimental study of the SP without need of going beyond critical Rabi coupling.
	
	\begin{figure}
		\includegraphics[width=1.0\columnwidth]{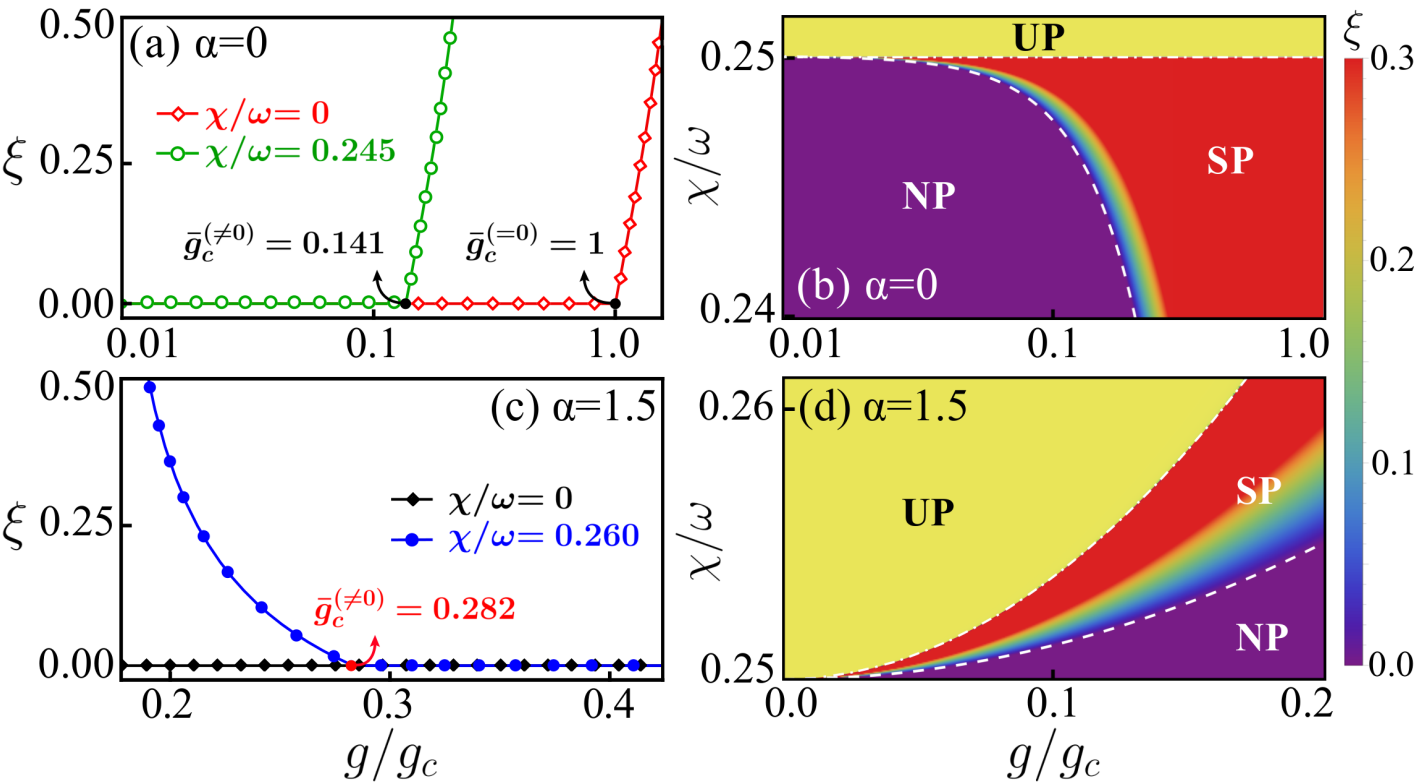}
		\caption{SPT at $\alpha=0$ (a, b) and $\alpha=1.5$ (c, d): The order parameter $\xi$ as functions of the coupling strength $g$ (a, c), phase diagram of $\xi$ in $g$-$\chi$ plane (b, d). The dashed (dot-dashed) line denotes the analytical boundary between SP and NP (UP).}
		\label{fig:phase_diagram}
	\end{figure}
	
In order to characterize the behavior of the SPT, we define the ground state photon number as the order parameter, i.e., $\xi = A\langle{\ada}\rangle_g$, with the scaling factor $A=e^{-2r}\omega /\Omega$ to unify different value cases of $\omega$ and $\chi$. In the limit of $\omega/\Omega\rightarrow 0$,  Hamiltonian~(\ref{eq:likeQRM}) can be readily diagonalized by expansion and unitary transformations~\cite{LiuM2017PRL,Ying-2021-AQT,Ying2020-nonlinear-bias,Lu-2018-1,Lu-2018-2}. Explicitly, $\xi = 0$ for $\overline{g} <  e^{-2r}$ and $\xi = \left(\tilde{\overline{g}}^2 - \tilde{\overline{g}}^{-2}\right)/4$ for $\overline{g}> e^{-2r}$. To show this clearly, we plot $\xi$ as a function of $\overline{g}$ in Fig.~\ref{fig:phase_diagram}(a, c). From Fig.~\ref{fig:phase_diagram}(a) where the AT is not included, the SPT can be induced no matter whether the magnon Kerr effect is taken into account or not. Without the magnon Kerr effect ($\chi=0$), the SPT occurs at $\overline{g}_c^{(=0)}=1$~(see the curve marked by diamonds). But when the magnon Kerr effect is considered ($\chi=0.245$), the SPT point is shifted to $\overline{g}_c^{(\neq0)}=0.141$~(curve marked by circles), which is much smaller than $\overline{g}_c^{(=0)}$, i.e., $\overline{g}_c^{(\neq0)}=0.141\overline{g}_c^{(=0)}$. This indicates that the introduced magnon Kerr effect can be utilized to dramatically reduce the critical coupling strength of the SPT, which greatly relaxes the parameter requirement in experiments. In Fig.~\ref{fig:phase_diagram}(c), the AT is considered ($\alpha=1.5$), one sees that the SPT disappears in the absence of the magnon Kerr effect~(black curve with solid diamonds). But when the magnon Kerr effect is introduced, the SPT is restored at $\overline{g}_c^{(\neq0)}=0.282$~(blue curve with dots). Note here, as afore-discussed, the SPT is reversed with the transition direction from SP to NP in coupling increasing, oppositely to the case in Fig.~\ref{fig:phase_diagram}(a).

	Fig.~\ref{fig:phase_diagram}(b, d) further shows ground-state phase diagram of $\xi$ in $g$-$\chi$ plane. In the absence of the AT but including the magnon Kerr effect~[Fig. \ref{fig:phase_diagram}(c) with $\alpha=0$], we can see that the transition from NP ($\xi=0$) to SP ($\xi>0$) can be observed in the increase of the coupling strength $g$ in $\chi/\omega < 1/4$ regime. The critical boundary is governed by $\chi /\omega=(1-\overline{g}^2)/4$~(dashed line). In $\chi/\omega>1/4$ regime, one has $\overline{g}^2={1- 4\chi/\omega}<0$ spuriously, the system enters an unstable phase~(UP) (yellow area). When both the AT and the magnon Kerr effect are included~[Fig.~\ref{fig:phase_diagram}(d) with $\alpha=1.5$], we find that both the SP and the NP can recover in the previously unstable regime of $\chi/\omega>1/4$, now with $1-4\chi/\omega$ and $(1-\alpha)$ both negative to fulfil $\overline{g}^2>0$. The critical NP/SP boundary is described by $\overline{g}=\sqrt{(1-4\chi/\omega)/(1-\alpha)}$~[dashed line in Fig.~\ref{fig:phase_diagram}(d)], while the SP/UP boundary is shifted from $\chi/\omega=1/4$ to $\chi/\omega=(1+\alpha \overline{g}^2)/4$~[dot-dashed line in Fig.~\ref{fig:phase_diagram}(d)].
	
	\begin{figure}
		\includegraphics[width=0.92\columnwidth]{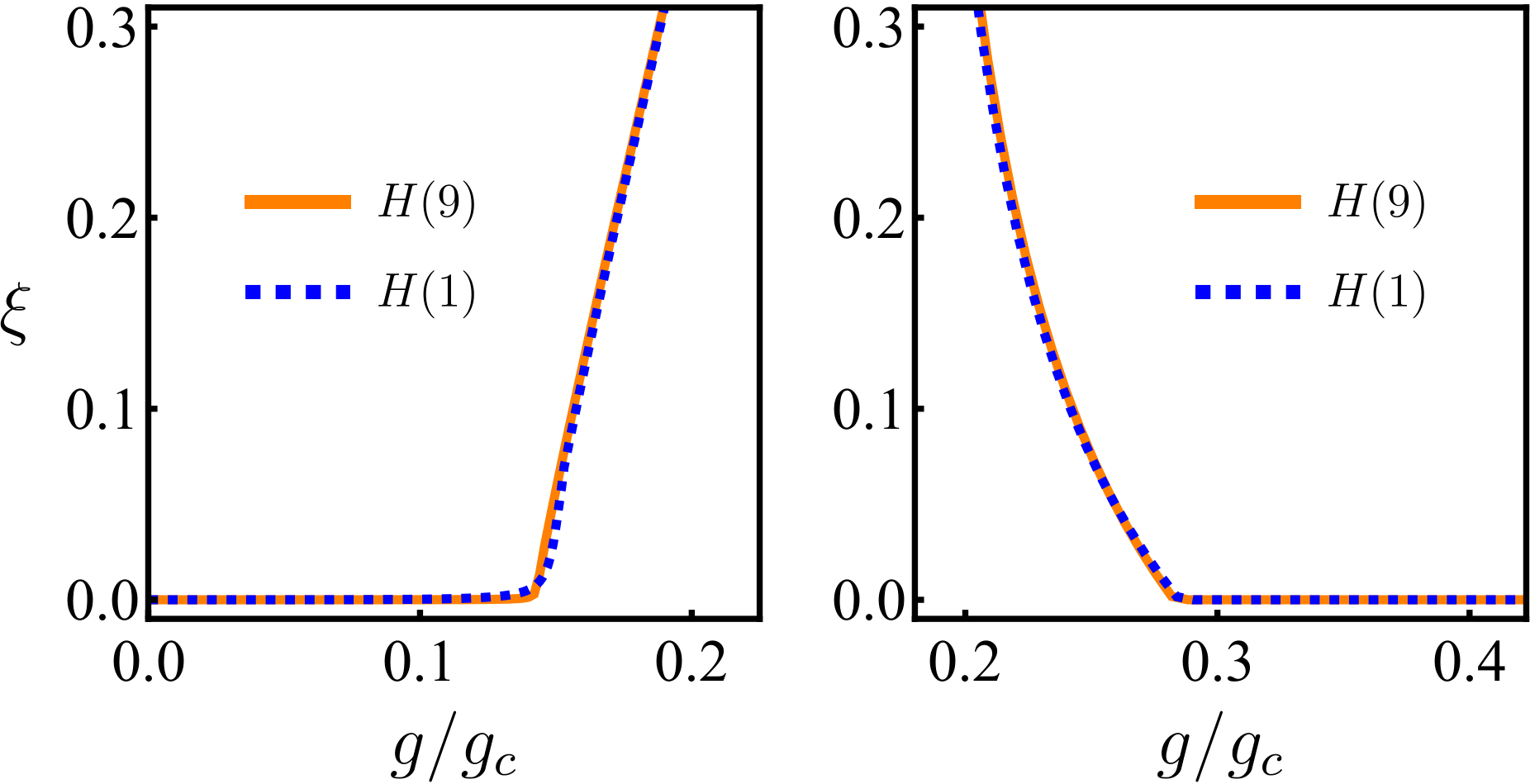}
		\caption{ A comparison of the order parameter $\xi$ as results of the effective Hamiltonian \eqref{eq9} (solid lines) and the original one \eqref{eq:totalHamiltonian} including the decay rates (dashed lines). (a) $\alpha=0$, with $K/\Omega =-6.125\times
10^{-13}$, $\omega _{m}/\Omega =367.5$; (b) $\alpha=1.5$, with $K/\Omega =-6.5\times 10^{-13}$, $\omega
_{m}/\Omega =390$. Here $\omega/\Omega = 0.01$, $\Omega =1$ MHz, $g_{m}/\Omega =2\pi \times 0.01$, $\kappa
_{m}/\Omega =2\pi \times 100,$ $\kappa _{a}/\Omega =2\pi \times 0.0001$ in both panels.}
		\label{fig:compare}
	\end{figure}

The results in Fig. \ref{fig:phase_diagram} are obtained from the effective
Hamiltonian in Eq.~\eqref{eq9}. Although Hamiltonian \eqref{eq9} is
analytically derived from the original Hamiltonian~%
\eqref{eq:totalHamiltonian} by adiabatically eliminating the magnon mode,
one may wonder about a numerical crosscheck. To confirm the validity of our
results, we further compare the order parameter $\xi $ with the result of the
original Hamiltonian \eqref{eq:totalHamiltonian}, as simulated by including
the decay rates of the cavity and the magnons via $\mathcal{H}_{\mathrm{eff}%
}=H-i\frac{\kappa _{m}}{2}m^{\dag }m-i\frac{\kappa _{a}}{2}a^{\dag }a$
according to the quantum Langevin equation. An example is illustrated in
Fig.~\ref{fig:compare} in the absence of the AT ($\alpha =0$ in panel (a)) and in the presence of the AT ($\alpha =1.5$ in panel (b)).
The parameters here meet the condition $\kappa _{m}\gg g_{m}$ as
previously applied in adiabatically eliminating the magnon mode. Here, in
both panels of Fig.~\ref{fig:compare}, the results of the effective
Hamiltonian~\eqref{eq9} and original Hamiltonian~\eqref{eq:totalHamiltonian}
are represented by the solid lines and dashed lines respectively. We see in
both cases, without the AT and with the AT, the results from the effective
Hamiltonian and original Hamiltonian are in good agreements, which shows that the mapping from Hamiltonian~\eqref{eq:totalHamiltonian} to
Hamiltonian~\eqref{eq9} is reliable and the SPT can be
indeed restored and reversed by the magnon Kerr effect, with the critical coupling strength significantly reduced.
	
	\section{Conclusion}
	
	In summary, we have proposed a hybrid quantum system combining nonlinear cavity magnonics and CQED to restore the SPT of the QRM, which has been thought to disappear in the presence of the AT due to the constraint of the no-go theorem, or to dramatically reduce the critical coupling strength if the SPT is not prohibited as in the counter no-go theorem in the debate. Indeed, by adiabatically eliminating the degrees of freedom of the magnons we have demonstrated that the Kerr magnons in a YIG sphere in coupling with the Rabi cavity system effectively introduce an additional AT which can counteract the intrinsic AT. The additional AT is not only tunable via the Kerr magnon effect but also can be strong as in nonlinear dependence of the magnon number which can be very large, thus being capable of making the SPT switchable. We have analytically extracted the critical coupling generally in the presence of both the AT and the Kerr effect. The recovered SPT is illustrated by the transition in the photon number and shown in an overall view by the figured-out phase diagram. We see that, when the AT is absent, our hybrid system can reduce the critical Rabi coupling thus greatly relaxes the experimental conditions for observing the SPT; When the intrinsic AT is included, the unreachable SPT without Kerr magnons can be gained by turning on the Kerr magnon-photon coupling, while the superradiant phase is available in small Rabi couplings due to the revered transition direction. We have also shown the magnonic parameter regimes for restoring the SPT which are experimentally tunable and accessible.
Note that an adjustable critical point is more favorable and provides more flexibility for applications such as in critical quantum metrology~\cite{Garbe2020,Garbe2021-Metrology,Ilias2022-Metrology,Ying2022-Metrology,Cai-2021,Wang-2018,Zhu-2019} where a wide range of critical couplings would much enlarge the critical sensing regime~\cite{Ying2022-Metrology}. In such a trend,
our proposal provides a promising path to manipulate the quantum phase transition with a hybrid system combing the advantages of nonlinear cavity magnonics and CQED.
		
	\section*{Acknowledgments}
	
	This work was supported by the National Natural Science Foundation of China (Grants No.~11974151 and No.~12247101) and the key program of the Natural Science Foundation of Anhui (Grant No. KJ2021A1301).

	
\end{document}